\documentclass[proceedings]{JHEP37}
\usepackage{eurosym}
\usepackage{amssymb}
\usepackage{amsfonts}
\usepackage{amsmath}
\usepackage{epsfig}

\newbox\mybox

\newcommand\fverb{\setbox\mybox=\hbox\bgroup\verb}
\newcommand\fverbdo{\egroup\medskip\noindent\fbox{\unhbox\mybox}\ }
\newcommand\fverbit{\egroup\item[\fbox{\unhbox\mybox}]}
\conference{Milne quantization for non-Hermitian systems}
\abstract{We generalize the Milne quantization condition to non-Hermitian systems. In the general case the underlying nonlinear Ermakov-Milne-Pinney equation needs to be replaced by a nonlinear integral differential equation. However, when the system is PT-symmetric or/and quasi/pseudo-Hermitian the equations simplify and one may employ the original energy integral to determine its quantization. We illustrate the working of the general framework with the Swanson model and two explicit examples for
pairs of supersymmetric Hamiltonians. In one case both partner Hamiltonians are Hermitian 
and in the other a Hermitian Hamiltonian is paired by a Darboux transformation to a non-Hermitian one.}

\title{Milne quantization for non-Hermitian systems}
\author{Sanjib Dey$^{1,2}$, Andreas Fring$^3$ and Laure Gouba$^4$ \\
$1$ Centre de Recherches Math\'{e}matiques (CRM), Universit\'{e} de Montr%
\'{e}al, \\
$\,\,$ Montr\'{e}al - H3C 3J7, Qu\'{e}bec, Canada\\
$2$ Department of Mathematics and Statistics, Concordia University, \\
$\,\,$ Montr\'{e}al - H3G 1M8, Qu\'{e}bec, Canada\\
$3$ Department of Mathematics, City University London, London EC1V 0HB, UK\\
$4$ The Abdus Salam International Centre for Theoretical Physics (ICTP),\\
$\,\,$ Strada Costiera 11, I-34151 Trieste Italy \\
E-mail: dey@crm.umontreal.ca,a.fring@city.ac.uk,lgouba@ictp.it}

\input{tcilatex}
\begin{document}

\section{Introduction}

As one of the first phase amplitude methods Milne provided in 1930 \cite%
{Milne} a relation between the time-independent Schr\"{o}dinger equation and
a non-linear integrable equation referred to these days as the
Ermakov-Milne-Pinney (EMP) equation \cite{Ermakov,Milne,Pinney} or variants
thereof. Solving either of the two equations for any generic energy will
provide a solution for the other. In addition, the interrelation involves an
auxiliary equation whose solutions lead to the exact energy quantization in
a very general fashion. It should be emphasized that the Milne quantization
is exact and the more popular WKB-approximation is obtained as a limiting
case when the second order derivative term in the EMP-equation is neglected.
While the latter method has been generalized \cite{benderwkb} to
non-Hermitian $\mathcal{PT}$-symmetric systems, this task has not been
carried out for the more general Milne quantization procedure. The main
purpose of this manuscript is to perform the first step in this direction
and to demonstrate that a successful application of the Milne quantization
procedure is indeed possible.

We analyze two types of non-Hermitian systems, in one case we exploit the
fact that the model is quasi/pseudo-Hermitian, the Swanson model, and in the
other that it is $\mathcal{PT}$-symmetric, a supersymmetric pair in which
one of the partner Hamiltonians is non-Hermitian.

Our manuscript is organized as follows: In section 2 we recall the key
features of the Milne quantization procedure and generalize it to a general
non-Hermitian setting. In section 3 we discuss the Swanson model and in
section 4 we provide two explicit examples for pairs of supersymmetric
Hamiltonians, where in one case both partner Hamiltonians are Hermitian and
in the other only one of them. Our conclusions and outlook are stated in
section 5.

\section{The Milne quantization for Hermitian and non-Hermitian systems}

We commence by briefly recalling the key idea of the solution procedure and
quantization method proposed originally by Milne in 1930 \cite{Milne}. Its
starting point is the time-independent Schr\"{o}dinger equation in the form%
\begin{equation}
\psi ^{\prime \prime }(x)+k^{2}(x)\psi (x)=0,  \label{Sch}
\end{equation}%
where the continuous energy parameter $E$ and the potential $V(x)$ are
combined into the local wavevector $k^{2}(x)=\hbar ^{2}/2m[E-V(x)]$.
Assuming the solution to equation (\ref{Sch}) to be of the general form%
\begin{equation}
\psi (x)=N\rho (x)\sin \left[ \phi (x)+\alpha \right] ,  \label{GS}
\end{equation}%
with normalization constant $N$, constant phase $\alpha $, amplitude $\rho
(x)$ and variable phase $\phi (x)$ a direct substitution leads to the
constraining equations%
\begin{equation}
\rho ^{\prime \prime }(x)+k^{2}(x)\rho (x)=\frac{\lambda ^{2}}{\rho ^{3}(x)}%
,\qquad \text{and\qquad }\rho ^{2}(x)\phi ^{\prime }(x)=\lambda ,
\label{EMP}
\end{equation}%
with $\lambda $ being some arbitrary constant. The first equation in (\ref%
{EMP}) is known as the Ermakov-Milne-Pinney (EMP) equation \cite%
{Ermakov,Milne,Pinney}. From (\ref{GS}) it is clear that its solution
together with a solution for the auxiliary equation for the phase function
will lead to an exact solution for the time-independent Schr\"{o}dinger
equation (\ref{Sch}) for generic values of $E$. Notice that when we neglect $%
\rho ^{\prime \prime }(x)$, the two equations in (\ref{EMP}) combine into $%
\phi ^{\prime }(x)=k(x)$ which corresponds to the WKB approximation. In what
follows we will employ Pinney's \cite{Pinney} general solution for the
EMP-equation\footnote{%
There exist other types of solutions, such as for instance the one reported
in \cite{eliezer} involving two free constants, which we may, however,
suitable chose.} 
\begin{equation}
\rho (x)=\sqrt{\psi _{1}^{2}(x)+\frac{\lambda ^{2}}{W^{2}}\psi _{2}^{2}(x)},
\label{SolEP}
\end{equation}%
with $\rho (x_{0})=\rho _{0}\neq 0$, $\rho ^{\prime }(x_{0})=\rho
_{0}^{\prime }$, $-\infty <x_{0}<\infty .$ Here $\psi _{1}$, $\psi _{2}$ are
the two fundamental solutions of the Schr\"{o}dinger equation (\ref{Sch})
and $W:=W(\psi _{1},\psi _{2})=\psi _{1}\psi _{2}^{\prime }-\psi
_{1}^{\prime }\psi _{2}$ denotes the corresponding Wronskian. Integrating
the second equation in (\ref{EMP}) directly and taking the initial
conditions to be $\psi _{1}(x_{0})=1$, $\psi _{2}(x_{0})=0$, $\psi
_{1}^{\prime }(x_{0})=1$, $\psi _{2}^{\prime }(x_{0})=\lambda $ implies $%
W=\lambda $ and leads to the general solution of equation (\ref{Sch})
expressed in terms of the solutions to the EMP-equation 
\begin{equation}
\psi (x)=N\rho (x)\sin \left[ W\int\nolimits_{x_{0}}^{x}\rho
^{-2}(s)ds+\alpha \right] .
\end{equation}%
Next we implement the boundary conditions. Demanding the wavefunction $\psi
(x)$ to vanish at the boundaries then implies the quantization condition 
\begin{equation}
I(E)=\frac{W(E)}{\pi }\int\nolimits_{-\infty }^{\infty }\rho
^{-2}(s,E)ds=n\in \mathbb{N}\text{,}  \label{EI}
\end{equation}%
when $\rho (x)$ is non-vanishing, meaning that any solution $E_{n}$ to $%
I(E_{n})=n$ constitutes a bound state energy. Note that the value of $I(E)$
is not sensitive to the normalization factors in the fundamental solution.

When the potential and possibly also the energy eigenvalues are complex the
general treatment is more involved. In that case we can make the Ansatz%
\begin{equation}
\psi (x)=N\rho (x)e^{i\phi (x)},  \label{Psi}
\end{equation}%
with $\rho (x),\phi (x)\in \mathbb{R}$ and separate the wavevector into its
real and imaginary part $k^{2}=\kappa +i\tau $. The substitution of (\ref%
{Psi}) into the time-independent Schr\"{o}dinger equation then yields the
two constraining equations when reading off the real and imaginary parts%
\begin{equation}
\rho ^{\prime \prime }(x)+\kappa (x)\rho (x)=\rho (x)\phi ^{\prime
}(x),\quad \text{and\quad }\phi ^{\prime \prime }(x)\rho (x)+2\phi ^{\prime
}(x)\rho ^{\prime }(x)+\tau (x)\rho (x)=0.
\end{equation}%
Combining these two equations generalizes the EMP-equation (\ref{EMP}) to 
\begin{equation}
\rho ^{\prime \prime }(x)+\kappa (x)\rho (x)=\frac{1}{\rho ^{3}(x)}\left(
\lambda -\int\nolimits^{x}\tau (s)\rho ^{2}(s)ds\right) ^{2}
\end{equation}%
with%
\begin{equation}
\phi (x)=\lambda \int\nolimits^{x}\rho ^{-2}(s)ds-\int\nolimits^{x}\rho
^{-2}(t)\left( \int\nolimits^{t}\tau (s)\rho ^{2}(s)ds\right) dt.
\end{equation}%
Evidently when $\tau =0$ we recover (\ref{EMP}). These equations are
difficult to solve, even in an approximate fashion. However, we may assume
that the quantization condition (\ref{EI}) still holds when $W(E)\in \mathbb{%
R}$ and $\limfunc{Im}[\rho ^{2}(s,E)]$ is an odd function in $s$. We will
demonstrate below that these properties can be attributed to the $\mathcal{PT%
}$-symmetry of the models.

\section{A quasi-Hermitian model, the Swanson Hamiltonian}

Quasi/pseudo-Hermitian Hamiltonian systems constitute a large subclass of
non-Hermitian systems \cite{Dieu,Urubu,Alirev}. They are characterized by
the fact that their non-Hermitian Hamiltonian $H$ can be mapped to an
isospectral Hermitian counterpart $h$ by means of a similarity
transformation $h=\eta H\eta ^{-1}$. The map $\eta $ is sometimes referred
to as the Dyson map \cite{Dyson} and satisfies certain properties. A prime
example for which this map and all other relevant quantities are known in
its explicit analytic form is the Swanson model \cite{Swanson}%
\begin{equation}
H_{S}=\omega \left( a^{\dagger }a+1/2\right) +\alpha a^{2}+\beta \left(
a^{\dagger }\right) ^{2},\qquad \omega ,\alpha ,\beta \in \mathbb{R},
\end{equation}%
with $a=\sqrt{\omega /2}x+i/\sqrt{2\omega }p$, $a^{\dagger }=\sqrt{\omega /2}%
x-i/\sqrt{2\omega }p$. Evidently $H_{S}$ is only Hermitian when $\alpha
=\beta $, but its isospectral Hermitian counterpart is known to be \cite{MGH}
\begin{equation}
h_{S}=\frac{\mu _{+}}{2}p^{2}+\frac{\mu _{-}}{2}x^{2},
\end{equation}%
with 
\begin{equation}
\mu _{\pm }=\frac{-\lambda (\alpha +\beta )+\omega \mp (\alpha +\beta
-\lambda \omega )\sqrt{1-\frac{(1-\lambda ^{2})(\alpha -\beta )^{2}}{(\alpha
+\beta -\lambda \omega )^{2}}}}{(1\pm \lambda )\omega ^{\pm 1}},\qquad
\lambda \in \lbrack -1,1].
\end{equation}%
The eigenvalue spectrum for both Hamiltonians is%
\begin{equation}
E_{n}=\left( n+\frac{1}{2}\right) \sqrt{\omega ^{2}-4\alpha \beta },\qquad
n\in \mathbb{N},
\end{equation}%
and thus real for $\omega ^{2}\geq 4\alpha \beta $. The corresponding
time-independent Schr\"{o}dinger equations are exactly solvable for both
Hamiltonians. The two fundamental solutions for the one corresponding to $%
h_{S}$ can be expressed in terms of parabolic cylinder functions, but in the
current context it is more convenient to employ the solutions in terms of
the closely related Whittaker functions 
\begin{eqnarray}
\psi _{1}(x) &=&\frac{1}{\sqrt{x}}M_{\frac{E}{2\sqrt{\mu _{-}\mu _{+}}},-%
\frac{1}{4}}\left( \sqrt{\frac{\mu _{-}}{\mu _{+}}}x^{2}\right) \Theta (x)+%
\frac{i}{\sqrt{x}}M_{\frac{E}{2\sqrt{\mu _{-}\mu _{+}}},-\frac{1}{4}}\left( 
\sqrt{\frac{\mu _{-}}{\mu _{+}}}x^{2}\right) \Theta (-x),\text{\quad } \\
\psi _{2}(x) &=&\frac{1}{\sqrt{x}}W_{\frac{E}{2\sqrt{\mu _{-}\mu _{+}}},-%
\frac{1}{4}}\left( \sqrt{\frac{\mu _{-}}{\mu _{+}}}x^{2}\right) \Theta (x)+%
\frac{i}{\sqrt{x}}W_{\frac{E}{2\sqrt{\mu _{-}\mu _{+}}},-\frac{1}{4}}\left( 
\sqrt{\frac{\mu _{-}}{\mu _{+}}}x^{2}\right) \Theta (-x).
\end{eqnarray}%
We neglect here normalization factors for the above mentioned reason. Unlike
the solutions in terms of parabolic cylinder functions this choice
guarantees that $\psi _{1,2}(x)\in \mathbb{R}$ or $\psi _{1,2}(x)\in i%
\mathbb{R}$, such that $\rho (x),W(E)\in \mathbb{R}$. Using these
expressions we compute the energy integral $I(E)$ in (\ref{EI}) and depict
our results in figure \ref{Swan}.

\FIGURE{\epsfig{file=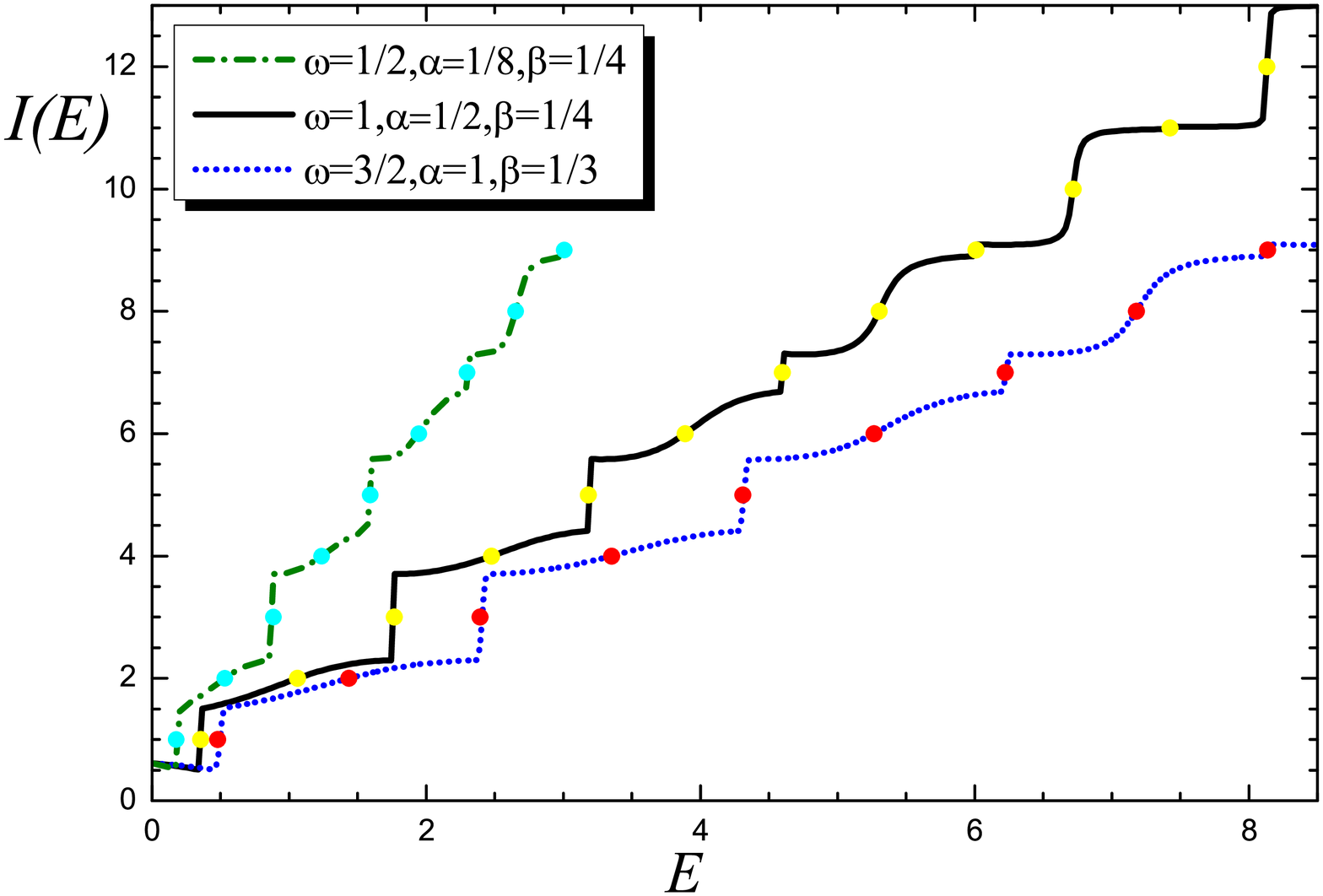,width=14.0cm}  
        \caption{Energy integrals $I(E)$ for the Swanson model, with 
$I((2n-1)/4\sqrt{2})=I(E_{n-1})=n \in \mathbb{N}$ for $\omega =1/2$, $\alpha =1/8$, $\beta =1/4$
   $I((2n-1)/2\sqrt{2})=I(E_{n-1})=n \in \mathbb{N}$ for $\omega =1$, $\alpha =1/2$, $\beta =1/4$ and 
$I((2n-1)11/4\sqrt{2})=I(E_{n-1})=n \in \mathbb{N}$ for $\omega =3/2$, $\alpha =1$, $\beta =1/3$}
        \label{Swan}}

The energy eigenvalues are located precisely at the expected values at
points of inflection of the function $I(E)$.

\section{Non-Hermitian models with supersymmetric Hermitian counterparts}

Now we study a model in which we exploit the $\mathcal{PT}$-symmetry of the
system. We consider a pair of supersymmetric quantum mechanical \cite%
{Witten:1981nf,Cooper:1982dm,Witten,Cooper} models described by the two
Hamiltonians%
\begin{equation}
H_{\pm }=L_{\pm }L_{\mp }=-\frac{d^{2}}{dx^{2}}+U^{2}(x)\pm U^{\prime }(x)=-%
\frac{d^{2}}{dx^{2}}+V_{\pm }(x),  \label{super}
\end{equation}%
involving the so-called superpotential $U(x)$. It is easily verified that
the solutions to the time-independent Schr\"{o}dinger equations $H_{\pm
}\psi _{\pm }=E\psi _{\pm }$ are related to each other by means of the two
intertwining operators $L_{\pm }$ 
\begin{equation}
L_{\pm }:=\pm \frac{d}{dx}+U(x),\qquad \quad \psi _{\pm }=\frac{1}{\sqrt{E}}%
L_{\pm }\psi _{\mp }.  \label{inter}
\end{equation}%
Denoting now the two fundamental solutions to the Schr\"{o}dinger equation
by $\psi $ and $\chi $, Ioffe and Korsch \cite{Ioffe:2002tk} found that the
corresponding Wronskians and solutions to the EMP-equations%
\begin{equation}
W_{\pm }:=W\left( \psi _{\pm },\chi _{\pm }\right) ,\qquad \rho _{\pm }=%
\sqrt{\psi _{\pm }^{2}+\chi _{\pm }^{2}}
\end{equation}%
are related to each other as%
\begin{equation}
W_{+}=W_{-},\qquad \text{and\qquad }E\rho _{\pm }^{2}=\left( L_{\pm }\rho
_{\mp }\right) ^{2}+\frac{W_{\mp }}{\rho _{\mp }^{2}}.  \label{IIK}
\end{equation}%
The first identity follows from a direct substitution of the wavefunction in
(\ref{inter}) into the defining relation for the Wronskian, the use of the
Schr\"{o}dinger equation and recalling that $dW/dx=0$. The derivation of the
second identity follows from a direct evaluation. We also add here for later
use an intermediate relation from that computation%
\begin{equation}
E\rho _{+}^{2}=U^{2}\rho _{-}^{2}+U(\rho _{-}^{2})^{\prime }+\left( \psi
_{-}^{\prime }\right) ^{2}+\left( \chi _{-}^{\prime }\right) ^{2}.
\label{I2}
\end{equation}

Let us now select our superpotentials to be of a very specific type, such
that one of the partner Hamiltonians is Hermitian whereas the other one is
not. Such a setting allows us to test our assertions from section 2. Bagchi
and Roychoudhury \cite{Bagchi:2000rr} provided a necessary condition for
such type of pairs and noted that one may even construct solvable models in
this case. Separating the real and imaginary parts in the superpotentials in
the form%
\begin{equation}
U(x)=a(x)+ib(x),\quad \text{with\quad }a(x),b(x)\in \mathbb{R}\text{, }a(x)=%
\frac{1}{2}\frac{d}{dx}\ln b(x),  \label{ab}
\end{equation}%
they observed that one obtains a real and a complex partner potential%
\begin{eqnarray}
V_{-}(x) &=&\frac{3b^{\prime 2}}{4b(x)^{2}}-\frac{b^{\prime \prime }(x)}{%
2b(x)}-b(x)^{2}\in \mathbb{R},\quad \text{\quad } \\
V_{+}(x) &=&\frac{b^{\prime \prime }(x)}{2b(x)}-\frac{b^{\prime 2}}{4b(x)^{2}%
}-b(x)^{2}+2ib^{\prime }(x)\notin \mathbb{R}.
\end{eqnarray}

In the following it will be important to utilize the effect of the parity
operator $\mathcal{P}$ and time-reversal operator $\mathcal{T}$ on the
various quantities involved. Our main requirement is that $V_{+}$ becomes $%
\mathcal{PT}$-symmetric, which is achieved as follows%
\begin{equation}
\mathcal{PT}:a(x)\rightarrow -a(x),b(x)\rightarrow b(x);\quad \mathcal{PT}%
:U(x)\rightarrow -U(x),V_{\pm }(x)\rightarrow V_{\pm }(x).
\end{equation}%
In order to obtain real eigenvalues $E\in \mathbb{R}$, usually referred to
as the spontaneously unbroken $\mathcal{PT}$-symmetric regime, we also
require the wavefunctions to be symmetric with regard to the anti-linear $%
\mathcal{PT}$-operator \cite{EW,Bender:1998ke} 
\begin{equation}
\mathcal{PT}:\psi _{\pm }(x),\rightarrow \psi _{\pm }(x),\chi _{\pm
}(x)\rightarrow \chi _{\pm }(x),W_{\pm }(x)\rightarrow -W_{\pm }(x),\rho
_{\pm }(x)\rightarrow \rho _{\pm }(x).
\end{equation}%
When assuming that $\psi _{-},\chi _{-}\in \mathbb{R}$, it follows from (\ref%
{I2}) and the subsequent use of the second relation in (\ref{IIK}) that%
\begin{equation}
\func{Im}\left( E\rho _{+}^{2}\right) =\func{Im}\left[ \left( L_{+}\rho
_{-}\right) ^{2}\right] =\frac{d}{dx}\left( b\rho _{-}^{2}\right) \text{.}
\end{equation}%
This implies that for real energies there will not be any contribution to
the integral in (\ref{EI}) from the imaginary part of the integrand $1/\rho
_{+}^{2}$ as it will be an odd function. The assumption $\psi _{-},\chi
_{-}\in \mathbb{R}$ also guarantees that $W_{-}\in \mathbb{R}$ and therefore
by the first relation in (\ref{IIK}) $W_{+}\in \mathbb{R}$, which are the
requirements mentioned at the end of section 2.

\subsection{A Hermitian/Hermitian supersymmetric pair}

As an illustration for the working of the conventional Milne quantization
for supersymmetric pairs we first consider a well studied exactly solvable
in the mathematical physics literature, \cite%
{Kleinert,Klauder2,dey_fring_pra}, the P\"{o}schl-Teller model \cite%
{Poschl:1933zz}. Taking the superpotential to be of the form%
\begin{equation}
U(x)=\lambda \tan x-\kappa \cot x,\qquad \kappa ,\lambda \in \mathbb{R}%
,0\leq x\leq \pi /2\text{.}
\end{equation}%
equation (\ref{super}) yields the pair of potentials%
\begin{equation}
V_{\pm }(x)=\lambda (\lambda \pm 1)\sec ^{2}x+\kappa (\kappa \pm 1)\csc
^{2}x-(\lambda +\kappa )^{2},
\end{equation}%
with $V_{-}(x)$ being the standard P\"{o}schl-Teller potential. The
fundamental solutions are well known. We have 
\begin{eqnarray}
\psi _{1}^{-}(x) &=&\sin ^{\kappa }x\cos ^{\lambda }x\,_{2}F_{1}\left[ \frac{%
\kappa +\lambda -\tilde{E}}{2},\frac{\kappa +\lambda +\tilde{E}}{2};\kappa +%
\frac{1}{2};\sin ^{2}x\right] , \\
\psi _{2}^{-}(x) &=&\sin ^{1-\kappa }x\cos ^{\lambda }x\,_{2}F_{1}\left[ 
\frac{1-\kappa +\lambda -\tilde{E}}{2},\frac{1-\kappa +\lambda +\tilde{E}}{2}%
;\frac{3}{2}-\kappa ;\sin ^{2}x\right] ,
\end{eqnarray}%
and%
\begin{eqnarray}
\psi _{1}^{+}(x) &=&\sin ^{\kappa +1}x\cos ^{\lambda +1}x\,_{2}F_{1}\left[ 
\frac{2+\kappa +\lambda -\tilde{E}}{2},\frac{2+\kappa +\lambda +\tilde{E}}{2}%
;\kappa +\frac{3}{2};\sin ^{2}x\right] , \\
\psi _{2}^{+}(x) &=&\sin ^{-\kappa }x\cos ^{\lambda +1}x\,_{2}F_{1}\left[ 
\frac{1-\kappa +\lambda -\tilde{E}}{2},\frac{1-\kappa +\lambda +\tilde{E}}{2}%
;\frac{1}{2}-\kappa ;\sin ^{2}x\right] ,
\end{eqnarray}%
where $_{2}F_{1}$ denoted hypergeometric function and we abbreviated $\tilde{%
E}:=\sqrt{(\kappa +\lambda )^{2}+E}$. Solutions to the EMP-equation are
simply obtained from (\ref{SolEP})%
\begin{equation}
\rho _{\pm }(x)=\sqrt{\left[ \psi _{1}^{\pm }(x)\right] ^{2}+\left[ \psi
_{2}^{\pm }(x)\right] ^{2}},
\end{equation}%
which allows us to compute the energy integrals (\ref{EI}) to 
\begin{equation}
I_{\pm }(E)=\frac{W_{\pm }(E)}{\pi }\int\nolimits_{0}^{\pi /2}\rho _{\pm
}^{-2}(s,E)ds.  \label{EI1}
\end{equation}%
Our numerical computations of (\ref{EI1}) are depicted in figure \ref{EnInt}.

\FIGURE{\epsfig{file=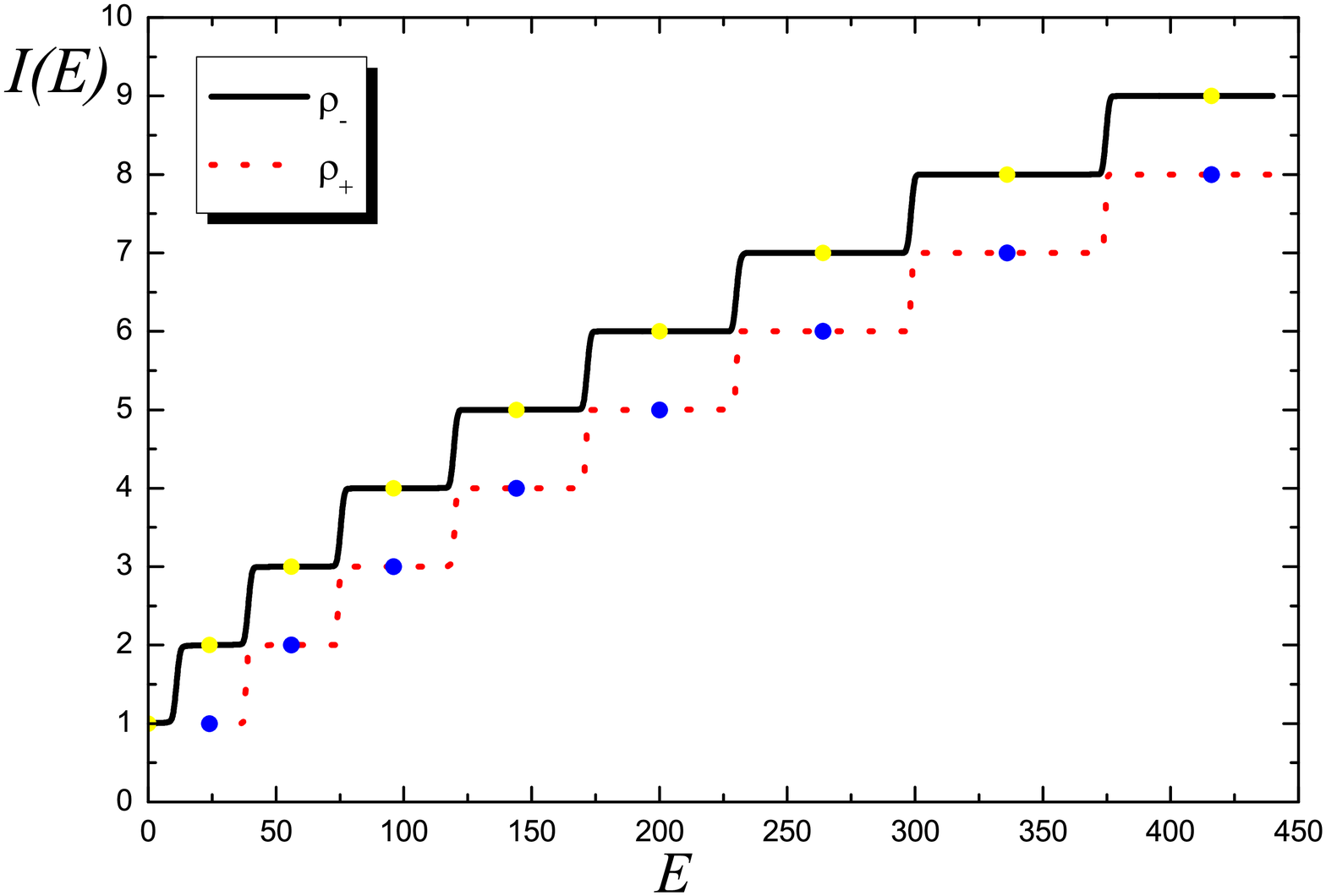,width=14.0cm}  
        \caption{Energy integrals $I_{\pm }(E)$ for a supersymmetric pair of P\"{o}schl-Teller potentials for coupling constants $\kappa =2$, $\lambda =3$, with $I_{-}(0)=I_{+}(24)=1$, $I_{-}(24)=I_{+}(56)=2$, $I_{-}(56)=I_{+}(96)=3$, $I_{-}(96)=I_{+}(144)=4$, $I_{-}(144)=I_{+}(200)=5$, $I_{-}(200)=I_{+}(264)=6$, $I_{-}(264)=I_{+}(336)=7$, $I_{-}(336)=I_{+}(416)=8
$ and $I_{-}(416)=9$.}
        \label{EnInt}}

For the selected values of the coupling constant $k=2$, $\lambda =3$ the
solutions to $I_{\pm }(E_{n}^{\pm })=n+1$ yield $E_{0}^{-}=0$, $%
E_{n}^{+}=E_{n+1}^{-}=4(n+1)(n+6)$ for $n=0,1,2,\ldots $ This is of course
the well known quantization condition obtained from demanding that $%
\lim_{x\rightarrow 0}\psi _{1}^{\pm }(x)=\lim_{x\rightarrow \pi /2}\psi
_{1}^{\pm }(x)=0$, achieved by setting the first entry of the hypergeometric
function $_{2}F_{1}$ to $-n$ with $n=0,1,2,\ldots $

\subsection{A Hermitian/Non-Hermitian supersymmetric pair}

Next we consider a superpotential giving rise to a Hermitian potential
paired with a non-Hermitian potential as proposed in \cite{Bagchi:2000rr}.
We take the superpotential $U(x)$ to be of the form 
\begin{equation}
U(x)=-\frac{1}{2}\tanh x+\frac{i}{2}(1-2\lambda )\func{sech}x,\qquad \lambda
\in \mathbb{R},
\end{equation}%
such that the real and imaginary parts are related as in (\ref{ab}). As
expected, when evaluating (\ref{super}) one of the partner potentials turn
out to be real%
\begin{equation}
V_{-}(x)=\frac{1}{4}+(\lambda -\lambda ^{2})\func{sech}^{2}x,  \label{Vmn}
\end{equation}%
whereas the other one becomes complex%
\begin{equation}
V_{+}(x)=\frac{1}{4}-\left( 1-\lambda +\lambda ^{2}\right) \func{sech}%
^{2}x+i(2\lambda -1)\func{sech}x\tanh x,  \label{Vpn}
\end{equation}%
albeit $\mathcal{PT}$-symmetric. The fundamental solutions are in this case%
\begin{eqnarray}
\psi _{1}^{-}(x) &=&\sinh x\cosh ^{\lambda }x\,_{2}F_{1}\left[ \mu _{-},\mu
_{+};\frac{3}{2};-\sinh ^{2}x\right] , \\
\psi _{2}^{-}(x) &=&\cosh ^{\lambda }x\,_{2}F_{1}\left[ \mu _{-}-\frac{1}{2}%
,\mu _{+}-\frac{1}{2};\frac{1}{2};-\sinh ^{2}x\right] ,
\end{eqnarray}%
and according to (\ref{inter}) we obtain the solutions for the partner
Hamiltonian as 
\begin{eqnarray}
\psi _{1}^{+}(x) &=&\frac{\cosh ^{\lambda -1}(x)}{12\sqrt{E}}\left[ 6\left[
2\cosh ^{2}x+(2\lambda -1)\sinh x(\sinh x-i)\right] \,_{2}F_{1}\left[ \mu
_{-},\mu _{+};\frac{3}{2};-\sinh ^{2}x\right] \right.   \notag \\
&&-\left. \frac{1}{4}\sinh ^{2}(2x)\left[ 4E+4\lambda (\lambda +2)+3\right]
\,_{2}F_{1}\left[ \mu _{-}+1,\mu _{+}+1;\frac{5}{2};-\sinh ^{2}x\right] %
\right] , \\
\psi _{2}^{+}(x) &=&\frac{\cosh ^{\lambda -1}(x)}{4\sqrt{E}}\left[
2(2\lambda -1)(\sinh x-i)\,_{2}F_{1}\left[ \mu _{-}-\frac{1}{2},\mu _{+}-%
\frac{1}{2};\frac{1}{2};-\sinh ^{2}x\right] \right.   \notag \\
&&+\left. \left( 1-4E-4\lambda ^{2}\right) \sinh x\cosh ^{2}x\,_{2}F_{1}
\left[ \mu _{-}+\frac{1}{2},\mu _{+}+\frac{1}{2};\frac{3}{2};-\sinh ^{2}x%
\right] \right] ,
\end{eqnarray}%
where $\mu _{\pm }:=(2+2\lambda \pm \sqrt{1-4E})/4$.

We have now all the ingredients to evaluate the energy integrals in (\ref%
{EI1}). Our results are depicted in figure \ref{EnInt2}.

\FIGURE{ \epsfig{file=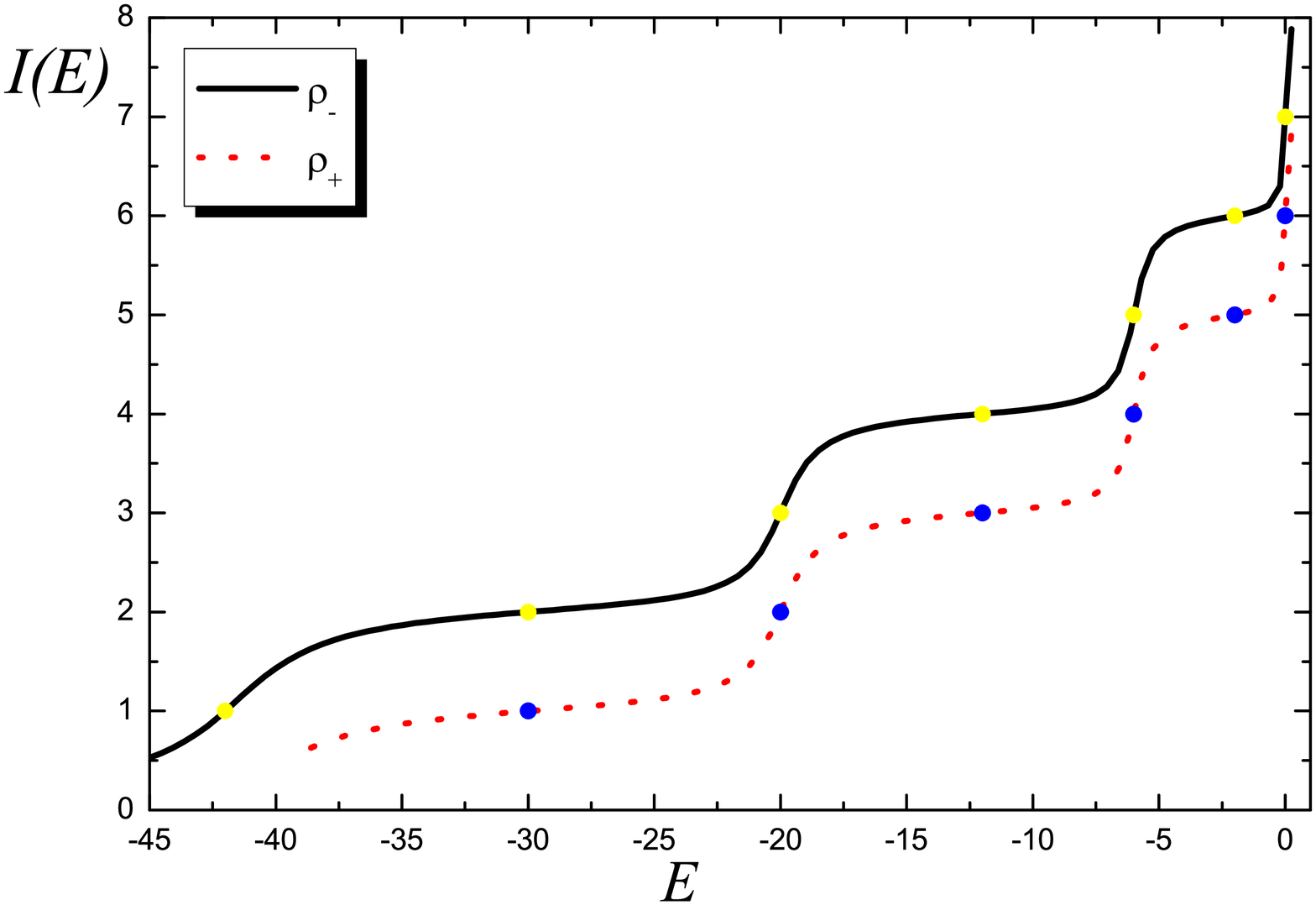,width=14.0cm} 
        \caption{Energy integrals $I_{\pm }(E)$ for a supersymmetric pair potentials $V_{\pm}$ in (\ref{Vmn}), (\ref{Vpn}) for the coupling constant $\lambda =15/2$, with $I_{-}(-42)=I_{+}(-30)=1$, $I_{-}(-30)=I_{+}(-20)=2$, $I_{-}(-20)=I_{+}(-12)=3$, $I_{-}(-12)=I_{+}(-6)=4$, $I_{-}(-6)=I_{+}(-2)=5$, $I_{-}(-2)=I_{+}(0)=6$, and $I_{-}(0)=7$.}
        \label{EnInt2}}

For the selected values of the coupling constant $\lambda $ the solutions to 
$I_{\pm }(E_{n}^{\pm })=n+1$ yield $E_{0}^{-}=-42$, $%
E_{n}^{+}=E_{n+1}^{-}=-(n-6)(n-7)$ for $n=0,1,2,\ldots $ This is again the
quantization condition obtained from demanding that $\lim_{x\rightarrow \pm
\infty }\psi _{1}^{\pm }(x)=0$, achieved by setting the first entry of the
hypergeometric function $_{2}F_{1}$ to $-n$ with $n=0,1,2,\ldots $The
remarkable feature is here that we can still use the standard formula for
the Milne quantization even though one of the Hamiltonians is non-Hermitian.
Notice that this feature can be attributed entirely to the $\mathcal{PT}$%
-symmetry of the system, which is responsible for the vanishing of the
imaginary part in the energy integral.

\section{Conclusion}

We demonstrated that the Milne quantization procedure can be successfully
adopted to non-Hermitian systems that are either quasi/pseudo-Hermitian or $%
\mathcal{PT}$-symmetric. For each scenario we provided an explicit example. We proposed some generalized formulae for the
generic non-Hermitian case, which are left as a challenge to be solved for
some concrete example.

Building on the success, it is to be expected that this method can be
applied also to systems for which the quantization is still incompletely
understood \cite{prep}, such as the complex Mathieu system currently of
great interest as it corresponds to the eigenvalue equation of the collision
operator in a two-dimensional Lorentz gas.

\medskip

\noindent \textbf{Acknowledgements}: SD is supported by the Postdoctoral Fellowship
jointly funded by the Laboratory of Mathematical Physics of the Centre de
Recherches Math\'{e}matiques (CRM) and by Prof. Syed Twareque Ali, Prof.
Marco Bertola and Prof. V\'{e}ronique Hussin. LG is supported by the Abdus
Salam International Centre for Theoretical Physics (ICTP).

\newif\ifabfull\abfulltrue

%%%\bibliographystyle{phreport}
%%%\bibliography{acompat,Ref}

\end{document}